\pdfoutput=1

\documentclass[11pt]{article}

\usepackage[]{naacl2021}

\usepackage{times}
\usepackage{latexsym}

\usepackage[T1]{fontenc}

\usepackage[utf8]{inputenc}

\usepackage{microtype}

\usepackage{graphicx}
\usepackage{multicol}
\usepackage{booktabs}
\usepackage{makecell}
\usepackage{amsmath}
\usepackage{multirow}
\usepackage{mathtools}
\usepackage{soul}
\usepackage{arydshln}
\usepackage{xcolor}

\title{Unsupervised Document Expansion for Information Retrieval\\ with Stochastic Text Generation}

\author{Soyeong Jeong$^1$
        \qquad Jinheon Baek$^2$
        \qquad ChaeHun Park$^1$
        \qquad Jong C. Park$^1$\thanks{\hspace{0.2cm}Corresponding author} \\
        School of Computing$^1$ \quad Graduate School of AI$^2$ \\
        Korea Advanced Institute of Science and Technology$^1$$^,$$^2$\\ 
       \texttt{\{syjeong,ddehun,park\}@nlp.kaist.ac.kr} \\
       \texttt{jinheon.baek@kaist.ac.kr}}

\begin{document}
\maketitle
\begin{abstract}

One of the challenges in information retrieval (IR) is the \textit{vocabulary mismatch} problem, which happens when the terms between queries and documents are lexically different but semantically similar. While recent work has proposed to expand the queries or documents by enriching their representations with additional relevant terms to address this challenge, they usually require a large volume of query-document pairs to train an expansion model. In this paper, we propose an \textbf{U}nsupervised \textbf{D}ocument \textbf{E}xpansion with \textbf{G}eneration (UDEG) framework with a pre-trained language model, which generates diverse supplementary sentences for the original document without using labels on query-document pairs for training. For generating sentences, we further stochastically perturb their embeddings to generate more diverse sentences for document expansion. We validate our framework on two standard IR benchmark datasets. The results show that our framework significantly outperforms relevant expansion baselines for IR\footnote{https://github.com/starsuzi/UDEG}.

\end{abstract}

\section{Introduction}

Information retrieval (IR) is the task of retrieving the most relevant documents, including scientific ones~\cite{Boudin2020Keyphrase, Noh_2020}, for a given query. IR systems have received considerable attention as they are not only required to search documents for information, but are also used as a core component in various downstream language understanding tasks such as open-domain question answering~\cite{Seo2019Realtime, Qu2020RocketQA}, fact verification~\cite{thorne-etal-2018-fact, Li2020Paragraph} and information extraction~\cite{Narasimhan2016Improving, das-etal-2020-information}. 

As the simplest approach to IR tasks, classical term-based ranking models, such as BM25~\cite{Robertson1994Okapi} and Query Likelihood (QL) models~\cite{Zhai2017Smoothing}, have been widely used. These term-based ranking models measure the lexical overlaps between query and document pairs using a sparse representation of words, to match the relevant documents for the given query. Notwithstanding their simplicity, they achieve decent performances, even compared to the recent dense representation models~\citep{Lin2019, xiong2020approximate}, which require a large number of paired query-document samples. However, these term-based sparse models are intrinsically vulnerable to the \emph{vocabulary mismatch} problem, which happens when a query and its relevant document are lexically divergent.

\begin{figure*}[t!]
\begin{center}
\includegraphics[width=0.7\textwidth]{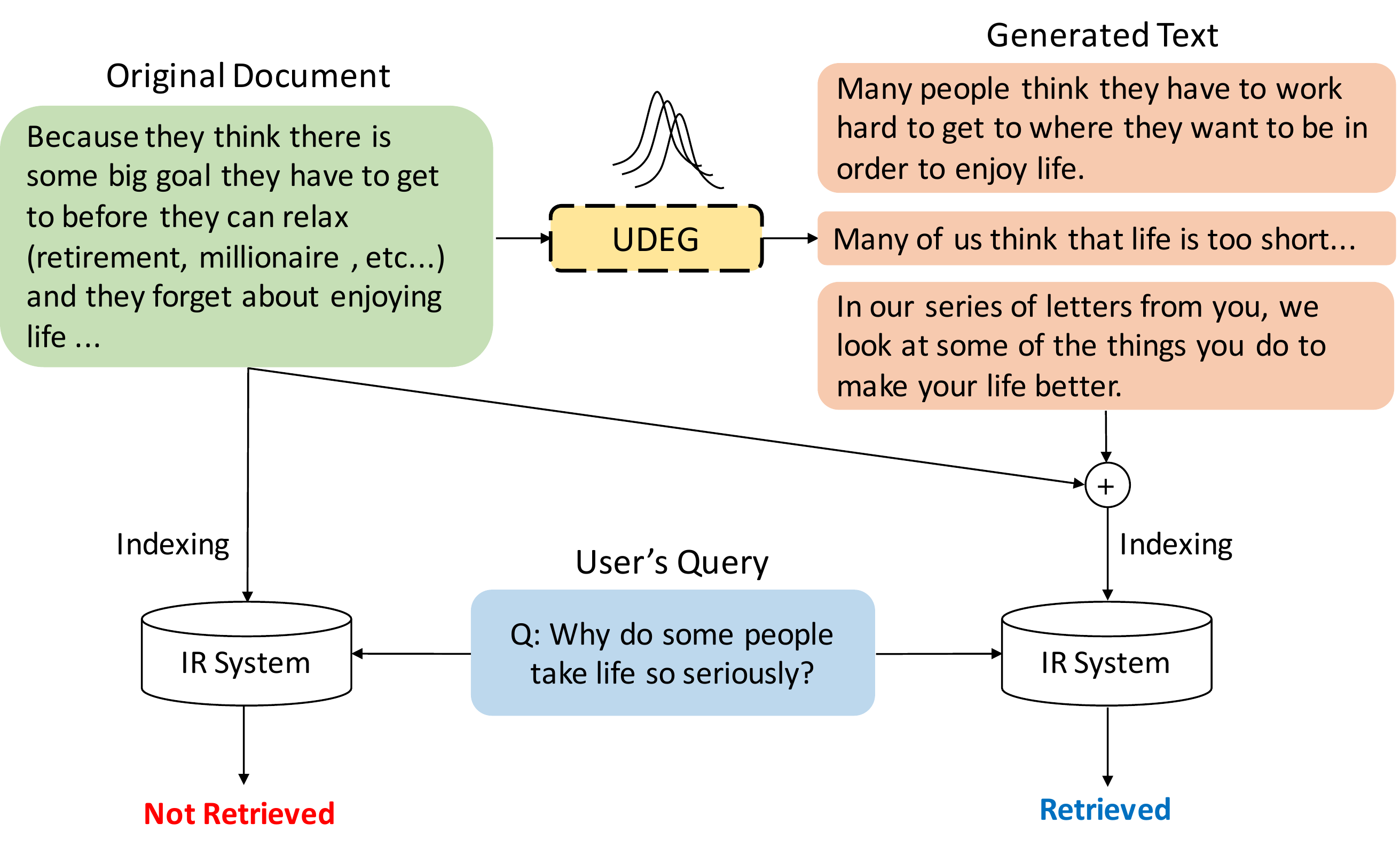}
\end{center}
\vspace{-0.15in}
\caption{The overall framework of our Unsupervised Document Expansion with Generation (UDEG), where the example is generated from our framework. Given an original document (green box), our UDEG framework stochastically generates several sentences (orange box) relevant to the given document, and augments the generated sentences to the input document to improve its expressiveness. After every document in the corpus is expanded, documents are indexed in the IR system, and searched in response to the given query.
}
\vspace{-0.1in}
\label{fig:fig1}
\end{figure*} 

Thus, we should address the limitations of both sparse and dense models, about the vocabulary mismatch problem and the need for a large amount of training data, respectively. Along this line, there are methods that expand queries and documents with their relevant terms. 
They include document expansion methods~\cite{Nogueira2019document, Boudin2020Keyphrase} that introduce additional context-related terms to given documents and query expansion methods~\cite{mao2020generationaugmented, Claveau2020Query} that augment given queries with additional terms. By doing so, we can explicitly generate lexically richer documents or queries.

Compared with query expansion, document expansion has two strengths. First, a document expansion model can generate much more relevant terms for the given document, since documents are generally much longer than queries. Also, documents can be expanded during indexing time so that the responding process for the user's query is not delayed, in contrast to queries that must be expanded during retrieval time. Thus, document expansion is more appropriate for a real-time system, together with making available more context-related words from the given information~\cite{Nogueira2019document}.

In this work, we focus on document expansion, and propose to abstractly generate the key information corresponding to the given document in an unsupervised manner, henceforth referred to as \emph{Unsupervised Document Expansion with Generation} (UDEG). We first generate document-related sentences using a pre-trained language model, and then stack up the newly generated sentences on the original documents to enrich the expressiveness of document representation. Specifically, in order to generate sentences containing particular information for the documents, we use a language model that is already trained for summarizing sentences from a sufficient amount of texts. However, such a scheme generates only one static sentence at a time, so we further propose to stochastically generate multiple relevant sentences for the given document. This helps the proposed UDEG framework to minimize the vocabulary mismatch cases by generating many relevant words, which reflect diverse points of view for the given document. The overall UDEG framework is illustrated in Figure~\ref{fig:fig1}.

We experimentally validate the proposed UDEG framework on standard benchmark datasets for IR tasks, ANTIQUE~\cite{Hashemi2020ANTIQUE} and MS MARCO~\cite{Nguyen2016MSMARCO}, with five different evaluation metrics. The experimental results show that our framework outperforms all baselines on all evaluation metrics by a large margin. Also, a detailed analysis of UDEG shows that its stochastic generation significantly improves the IR performances, and that our UDEG framework does not depend on specific language models for generation.

Our contributions in this work are threefold: 
\begin{itemize}
  \item To mitigate the vocabulary mismatch problem, we present a novel document expansion framework that augments the document with abstractly generated sentences without using paired query-document data for training.  
  \item Under an unsupervised document expansion framework, we generate document-related sentences with a pre-trained language model, and further stochastically perturb the embeddings for more diverse sentences.
  \item We show that our framework achieves outstanding performances on benchmark datasets for IR tasks with various evaluation metrics.
\end{itemize}
\section{Related work}

\paragraph{Information Retrieval}
A two-stage pipeline is the most prominent approach for IR. This pipeline first retrieves query-relevant documents with their sparse representations, and then re-ranks them by using neural networks~\cite{Mitra2018An, Nogueira2019document, Nogueira2020DocumentRanking}. 
In this two-stage pipeline, the overall performance is critically dependent on the first retrieval stage, since the failure of the retrieval stage would highly affect the second re-ranking stage. 
Therefore, this bottleneck on the first stage has to be addressed for performance enhancement~\cite{Karpukhin2020Dense}. BM25 and query likelihood (QL) are the most popular \textit{ad-hoc} retrieval models for the first stage~\cite{Nogueira2019document, Boudin2020Keyphrase, tang2020neural}. More recently, instead of using sparse models, methods of using dense representations have been proposed~\cite{Karpukhin2020Dense, xiong2020approximate, Qu2020RocketQA}, which can help alleviate the vocabulary mismatch problem through a dense representation space. 
However, recent work has revealed the limitations on their performance and efficiency~\cite{Lin2019, xiong2020approximate, Luan2020sparse}. 
Furthermore, these dense representation methods are based on supervised learning, where pairs of query and related-document are usually required to ensure reasonable performance.

\paragraph{Query / Document Expansion} 
Query and document expansions have been widely used in IR systems. 
In terms of query expansion, \citet{Nasreen2004UMass} proposed pseudo relevance feedback (RM3), which is revisited in more recent work~\cite{2020DibiaNeuralQA, mao2020generationaugmented} for its strength. There are also methods that expand queries using generation schemes~\cite{mao2020generationaugmented, Claveau2020Query}.
However, query expansion suffers from its intrinsic drawbacks, as queries need to be manipulated during the retrieval phase and have relatively less information than documents~\cite{Nogueira2019document}.
Thus, we take the alternative route: expanding documents. 
\citet{Nogueira2019document} and \citet{tang2020neural} proposed to expand documents with generated text using a supervised model trained on query-document pairs.
In contrast, our framework generates document-related sentences regardless of the existence of the corresponding query. 
\citet{Boudin2020Keyphrase} proposed to expand documents with sequence-to-sequence models which output keyphrases; however, their models have to be trained from scratch on a specific domain.

\paragraph{Document-relevant Text Generation}
In order to enrich the given document efficiently with the document-relevant text, such text should contain the document's key context which can appear in the summarized sentence. 
Earlier, \citet{Erkan2004lexrank} and \citet{mihalcea-tarau-2004-textrank} proposed unsupervised models of extracting key sentences, which are adopted in various recent work for their robustness~\cite{ Nikola2020Abstractive, Zhang2020Keywords, kazemi-etal-2020-biased}. 
In contrast, an abstractive approach aims at generating summarized sentences containing novel terms that might not exist in the given document~\cite{Zhang2020PEGASUS, Yang2020TED}.
~\citet{Nikolov2020Abstractive} proposed to first extract the key sentences and then paraphrase them with back-translation. 
Recent work has reported that the improved performance of text summarization approaches is attributed to the pre-trained language models~\cite{Zhang2020PEGASUS, Lewis2020BART, Xu2020Understanding}.
In this work, we aim at abstractly generating a document-related sentence with a pre-trained language model and further propose to diversely generate sentences with stochastic perturbation, not just using a single summarized sentence.
\section{Method}
Our goal is to expand the document for IR tasks by generating document-related text, which contains novel but semantically similar terms for the given document without using query-document pairs. In this section, we describe formal description of the IR task.

\subsection{Preliminaries}
We begin with a formal description of the IR task, and then introduce a document expansion scheme.

\paragraph{Information Retrieval}

The objective of an IR task is to retrieve the most relevant document $\boldsymbol{d} \in \mathcal{D}$ for the given query $\boldsymbol{q} \in \mathcal{Q}$, where $\mathcal{Q}$ and $\mathcal{D}$ indicate query and document set, respectively. Note that the query and document pair can be represented as either sparse~\cite{Robertson1994Okapi, Zhai2017Smoothing} or dense~\citep{Lin2019, xiong2020approximate}, which gives rise to different implementation details.

Suppose that we are given a query-document pair $(\boldsymbol{q}, \boldsymbol{d})$ in the correct query-document set $\tau$: $(\boldsymbol{q}, \boldsymbol{d}) \in \tau$, where $\tau \subset \mathcal{Q} \times \mathcal{D}$. Then, the system should retrieve the most relevant document $\boldsymbol{d}$ for the given query $\boldsymbol{q}$ in the correct query-document set $\tau$, denoted as follows:
\vspace{-0.05in}
\begin{equation}
    \max \sum_{(\boldsymbol{q}, \boldsymbol{d}) \in \tau} f(\boldsymbol{q}, \boldsymbol{d}),
\label{eq:ir:objective}
\end{equation}
where $f: \mathcal{Q} \times \mathcal{D} \rightarrow \mathcal{R}$ is a score function that measures the similarity of the correct query-document pairs, to retrieve the most relevant document for the given but unseen query at test time.
\vspace{-0.05in}

\paragraph{Document Expansion}
While an IR system can work alone as in Equation~\ref{eq:ir:objective} by using either sparse or dense representations for queries and documents, we need to deal with the vocabulary mismatch problem, which happens when the terms between queries and documents are lexically different but semantically related. To address this problem, we focus on the document expansion scheme, which augments the document with relevant terms to make a richer document.

Formally, the goal of document expansion is to generate semantically relevant terms $\boldsymbol{t} = \left[ t_i \right]_{i=1}^{K}$ for the given document $\boldsymbol{d} \in \mathcal{D}$, denoted as follows:
\vspace{-0.05in}
\begin{equation}
    \left[ t_i \right]_{i=1}^{K} = g(\boldsymbol{d}; \boldsymbol{\theta}),
\label{eq:term}
\end{equation}
where $K$ is the number of terms associated with each document $\boldsymbol{d}$ and $g$ is the document expansion model parameterized by $\boldsymbol{\theta}$. After generating relevant terms $\boldsymbol{t} = \left[ t_i \right]_{i=1}^{K}$ for the document, we concatenate them with the original document $\boldsymbol{d}$ to construct the more meaningful document-representation $\boldsymbol{\bar{d}}$, denoted as follows:
\vspace{-0.05in}
\begin{equation}
    \boldsymbol{\bar{d}} = \left[ \boldsymbol{t} \oplus \boldsymbol{d} \right],
\label{eq:concat}
\end{equation}
where $\oplus$ is the concatenation operation.

By expanding relevant terms to the given document with the generation model $g$, the similarity between the query $\boldsymbol{q}$ and expanded document $\boldsymbol{\bar{d}}$ becomes stronger than the similarity between query $\boldsymbol{q}$ and original document $\boldsymbol{d}$, as follows: $f(\boldsymbol{q}, \boldsymbol{d}) \le f(\boldsymbol{q}, \boldsymbol{\bar{d}})$. 
In order to maximize the similarity score between $\boldsymbol{q}$ and $\boldsymbol{\bar{d}}$, we need the model $g$ that generates document-related terms without using labels of query-document pairs $\tau$ for training, which we describe in the next subsection.

\begin{table}[t]
\small
\centering

\begin{tabular}{ll}
\toprule

\textbf{Document (Input):} Because they think there is some big \\
goal they have to get to before they can relax (retirement, \\
millionaire, etc...) and they forget about enjoying life ...  \\

\toprule

\textbf{Ext (Output):} relax (retirement, millionaire , etc...) \\

\cdashline{0-0}\noalign{\vskip 0.75ex}

\textbf{Abs (Output):} Many people think they have to work \\
hard to get to where they want to be in order to enjoy life. \\

\cdashline{0-0}\noalign{\vskip 0.75ex}

\textbf{Abs + S (Output):} \textbf{1)} Many people think they have to \\
work hard to get to where they want to be in order to \\
enjoy life. \\
\textbf{2)} Many of us think that life is too short... \\
\textbf{3)} In our series of letters from you, we look at some of \\
the things you do to make your life better. \\

\bottomrule
\end{tabular}
\vspace{-0.10in}
\caption{Examples of the generated text from different text-generation schemes. Generated terms are appended to the input document before indexing them for the IR system. Ext and Abs denote the extractive and abstractive generation, respectively. Also, + S symbol denotes the stochastic generation.}
\label{tbl:method:example}
\vspace{-0.20in}
\end{table}

\subsection{Unsupervised Text Generation for Document Expansion}
We now describe our \emph{Unsupervised Document Expansion with Generation} (UDEG) framework, which generates relevant terms for the given document $\boldsymbol{d}$ without using labels on query-document pairs $(\boldsymbol{q}, \boldsymbol{d}) \in \tau$. We first introduce the extractive and abstractive text generation schemes, which are two representative methods for unsupervised text generation, and then propose a stochastic generation scheme for a richer vocabulary.

\paragraph{Extractive Text Generation}
Extractive text generation is to select the representative words or sentences on the given document. Formally, an extractive text generation scheme is defined as follows:
\vspace{-0.05in}
\begin{equation}
\begin{gathered}
    \boldsymbol{t}_{ext} = \left[ (t_{ext})_i \right]_{i=1}^{K} = g_{ext}(\boldsymbol{d}; \boldsymbol{\theta}_{ext}), \\ \left[ (t_{ext})_i \right]_{i=1}^{K} \subset \boldsymbol{d}, 
\end{gathered}
\label{eq:ext}
\end{equation}
where $g_{ext}$ is an extractive text generation model parameterized by~$\boldsymbol{\theta}_{ext}$. After extracting terms $\boldsymbol{t}_{ext} = \left[ (t_{ext})_i \right]$, they are used to expand the document as in Equation~\ref{eq:concat} (i.e., $\boldsymbol{\bar{d}} = \left[ \boldsymbol{t}_{ext} \oplus \boldsymbol{d} \right]$), which can enrich the representation of the given document by counting important terms multiple times (See Table~\ref{tbl:method:example} for examples of extractive generation).

\paragraph{Abstractive Text Generation}
While the previously described extractive text generation model aims at enriching the given document with key terms extracted from it, the expressiveness of this extractive scheme is highly restricted since novel but semantically similar terms cannot be generated as in Equation~\ref{eq:ext}: $\left[ (t_{ext})_i \right]_{i=1}^{K} \subset \boldsymbol{d}$. To overcome this limitation, one should further consider generating related-terms that are not contained in the original document. To this end, we propose an abstractive text generation model to obtain the relevant but novel terms for the given document $\boldsymbol{d}$.

Formally, novel terms for the original document are denoted as $\left[ (t'_{abs})_l \right]_{l=1}^{N} \not\subset \boldsymbol{d}$, whereas existing terms on the document are denoted as $\left[ (t_{abs})_j \right]_{j=1}^{K-N} \subset \boldsymbol{d}$. $N$ is the number of newly generated document-related terms. Then, an abstractive generation model is defined as follows:
\vspace{-0.05in}
\begin{equation}
\begin{gathered}
    \boldsymbol{t}_{abs} = \left[ \left[ (t'_{abs})_l \right]_{l=1}^{N} \oplus \left[ (t_{abs})_j \right]_{j=1}^{K-N} \right]  \\
    = g_{abs}(\boldsymbol{d}; \boldsymbol{\theta}_{abs}), 
\label{eq:abs}
\end{gathered}
\end{equation}
where $g_{abs}$ is the abstractive generation model parameterized by~$\boldsymbol{\theta}_{abs}$. We provide concrete examples of abstractive generation in Table~\ref{tbl:method:example}.

Specific details of unsupervised text generation models, which do not use labels for query-document pairs, are described in \S~\ref{sub4.2}.

\paragraph{Stochastic Generation}
While a naïve abstractive generation scheme can generate novel terms that are not included in the original document, a major drawback of this scheme is that they cannot generate a high volume of different terms for the given document. 
In other words, this scheme is suboptimal since it only generates a single sequence, though the terms within the document can have many synonymous expressions.
To overcome this limitation, we stochastically generate terms for the given document by perturbing its embeddings for text generation via applying Monte Carlo~(MC) dropout~\citep{mcdropout}. Compared to the abstractive generation scheme in Equation~\ref{eq:abs}, which only produces one typical sequence of terms~$\boldsymbol{t}_{abs}$, we obtain $S$ different sequences $\boldsymbol{T}_{abs}$ from the stochastic generation scheme, as follows:
\begin{equation}
\begin{gathered}
    \boldsymbol{T}_{abs} = \left[ \boldsymbol{t}_{abs}^{i} \right]_{i=1}^{S} \\ \boldsymbol{t}_{abs}^{i} = g_{abs}'(\boldsymbol{d}; \boldsymbol{\theta}_{abs}),
\end{gathered}
\end{equation}
where $g_{abs}'$ randomly masks weights on the model even at test time. We provide examples of stochastic generation with $S=3$ in Table~\ref{tbl:method:example}. 
As shown in Table~\ref{tbl:method:example}, examples of stochastic generation are more relevant to the document and more diverse.
\section{Experimental Setups}
\aboverulesep=0ex
\belowrulesep=0ex

\begin{table*}
\centering
\begin{center}
\resizebox{0.9\width}{!}{
\begin{tabular}[t]{ll|cccccc}
\toprule
\multicolumn{2}{c|}{} & \textbf{No Expan.} & \textbf{MP-rank} & \textbf{LexRank} &  \textbf{Lex.+Para.} & \textbf{PEGASUS$_{ext}$} & \textbf{UDEG (Ours)} \\
\midrule
\multirow{4}{*}{MRR} 
& BM25 & 0.595 & 0.584 & 0.571 & 0.561 & 0.585 &\textbf{0.645}\\
& BM25+RM3 & 0.558 & 0.579 & 0.542 & 0.567 & 0.555 &\textbf{0.616}\\
& QL & 0.499 & 0.534 & 0.567 & 0.518 & 0.562 &\textbf{0.650}\\
& QL+RM3 & 0.396 & 0.447 & 0.456 & 0.432 & 0.504 &\textbf{0.583}\\
\hline
\multirow{4}{*}{R@10} 
& BM25 & 0.218 & 0.220 & 0.208 & 0.209 & 0.207 &\textbf{0.237}\\
& BM25+RM3 & 0.217 & 0.221 & 0.208 & 0.204 & 0.213 &\textbf{0.226}\\
& QL & 0.189 & 0.199 & 0.203 & 0.196 & 0.205 &\textbf{0.232}\\
& QL+RM3 & 0.159 & 0.179 & 0.182 & 0.162 & 0.191 &\textbf{0.211}\\
\hline
\multirow{4}{*}{P@3} 
& BM25 & 0.378 & 0.381 & 0.346 & 0.351 & 0.356 &\textbf{0.431}\\
& BM25+RM3 & 0.361 & 0.355 & 0.360 & 0.373 & 0.366 &\textbf{0.433}\\
& QL & 0.301 & 0.333 & 0.340 & 0.315 & 0.358 &\textbf{0.418}\\
& QL+RM3 & 0.240 & 0.281 & 0.275 & 0.271 & 0.301 &\textbf{0.386}\\
\hline
\multirow{4}{*}{MAP} 
& BM25 & 0.211 & 0.212 & 0.199 & 0.202 & 0.201 &\textbf{0.238}\\
& BM25+RM3 & 0.212 & 0.213 & 0.203 & 0.203 & 0.207 &\textbf{0.234}\\
& QL & 0.172 & 0.191 & 0.192 & 0.181 & 0.199 &\textbf{0.230}\\
& QL+RM3 & 0.150 & 0.168 & 0.170 & 0.158 & 0.180 &\textbf{0.212}\\
\hline
\multirow{4}{*}{NDCG@3} 
& BM25 & 0.437 & 0.442 & 0.417 & 0.425 & 0.419 &\textbf{0.478}\\
& BM25+RM3 & 0.424 & 0.434 & 0.423 & 0.433 & 0.426 &\textbf{0.470}\\
& QL & 0.356 & 0.389 & 0.400 & 0.375 & 0.413 &\textbf{0.471}\\
& QL+RM3 & 0.277 & 0.324 & 0.319 & 0.306 & 0.350 &\textbf{0.424}\\
\bottomrule
\end{tabular}
}
\end{center}
\vspace{-0.15in}
\caption{Retrieval results on the ANTIQUE dataset. We use five evaluation metrics: MRR, R@10, P@3, MAP, and NDCG@3. Also, the best performance is marked in \textbf{bold}.}
\label{tab:tab1}
\vspace{-0.15in}
\end{table*}

Here, we describe datasets, models, evaluation metrics, and implementation details for experiments.

\subsection{Datasts}
We use two benchmark datasets for IR to evaluate our UDEG framework as follows:

\noindent\textbf{ANTIQUE:}
This is a dataset with 403,666 documents from Yahoo! Answer, including open-domain non-factoid questions~\cite{Hashemi2020ANTIQUE}. The test set consists of 200 queries and 6,589 query-document pairs.

\noindent\textbf{MS MARCO:}
This is a collection of 8,841,823 passages from Bing search engine~\cite{Nguyen2016MSMARCO}. Since the test set is not publicly available, we use the development set containing 6,980 queries and 59,273 query-document pairs. We randomly sample 1,000,000 passages, while using the same development set for queries and query-document pairs, due to the limitation of computational resources on expanding 8,841,823 passages.

\subsection{Retrieval Models}
In this subsection, we describe two retrieval models that are widely used for IR systems.

\noindent\textbf{BM25:}
This is one of the standard \textit{ad-hoc} retrieval models based on Term Frequency-Inverse Document Frequency (TF-IDF), which measures overlapping terms between query and document~\cite{Robertson1994Okapi}.

\noindent\textbf{QL:}
This is also one of the standard \textit{ad-hoc} retrieval models. Specifically, QL returns a ranked list of documents sorted by the probability of $P(d|q)$, where $q$ is a query and $d$ is a document~\cite{Zhai2017Smoothing}.

\subsection{Expansion Models}
We compare our UDEG framework against the following baselines: 

\noindent\textbf{No Expansion (No Expan.):}
This is a naïve model of retrieving the original documents without query or document expansion. 

\noindent\textbf{RM3:}
This is a query expansion model that uses a pseudo-relevance feedback scheme (RM3)~\cite{Nasreen2004UMass}. Note that this can be simultaneously used with document expansion models.

\noindent\textbf{MP-rank:}
This is an extractive document expansion model, which extracts keyphrases based on a multipartite graph, where the nodes are keyphrase candidates and an edge connects nodes having different topics~\cite{Boudin2018Unsupervised}.

\noindent\textbf{LexRank:}
This is an extractive document expansion model that extracts the key sentence with PageRank algorithm~\cite{Page1998pagerank}, which constructs vertices as sentences and edges as TF-IDF weights~\cite{Erkan2004lexrank}.

\noindent\textbf{PEGASUS$_{ext}$:}
This is an extractive document expansion model~\cite{Zhang2020PEGASUS}, which extracts sentences using pre-trained knowledge for generating masked sentences on the CNN/DailyMail dataset~\cite{nallapati-etal-2016-abstractive}.

\noindent\textbf{LexRank + paraphrase (Lex. + Para.):}
This is an abstractive document expansion model, which first extracts key sentences with LexRank, and then paraphrases them with an unsupervised model~\cite{Liu2020UPSA} based on simulated annealing.

\noindent\textbf{UDEG: }
Our framework of expanding documents with abstractly generated sentences from a pre-trained language model. Diverse sentences are generated with stochastic perturbation by MC dropout.
\label{sub4.2}

\subsection{Metrics}
We evaluate the models with five metrics, ranging from precision- to recall-oriented, as follows: 

\noindent\textbf{Mean Reciprocal Rank (MRR):}
MRR measures the location of the first relevant document for the given query in a binary sense.

\noindent\textbf{Recall (R@K):}
R@K measures the recall up to K recommended documents.

\noindent\textbf{Precision (P@K):}
P@K measures the precision up to K recommended documents. 

\noindent\textbf{Mean Average Precision (MAP):}
Similar to P@K, MAP evaluates all related documents with an ordered list of them.

\noindent\textbf{Normalized Discounted Cumulative Gain (NDCG@K):}
Compared to the MAP that uses binary relevance metrics, this further manipulates the recommended list by using the fact that some documents are more relevant than others.

\subsection{Implementation Details}

All of the retrieval models are implemented using Anserini open-source IR toolkit~\cite{Yang2018anserini} with the default hyperparameter values. The PEGASUS-large model, already fine-tuned on the XSUM dataset~\cite{Narayan2018xsum}, is used as a pre-trained language model in UDEG for abstractive text generation. 
For the decoding algorithm, we use a beam search algorithm and set the beam size as 8. Also, we set the number $S$ of stochastic generation for document expansion as 4.


\section{Results and Discussion}
In this section, we show the overall performance of our UDEG, and then analyze the results in detail. 
\label{sec5}

\subsection{Overall Results}
\aboverulesep=0ex
\belowrulesep=0ex

\begin{table}
\centering
\begin{center}
\resizebox{0.48\textwidth}{!}{
\begin{tabular}[t]{ll|ccccc}
\Xhline{2\arrayrulewidth}
\toprule
\multicolumn{2}{c|}{} & \textbf{No Expan.} & \textbf{LexRank} &  \textbf{UDEG (Ours)} \\
\midrule
\multirow{4}{*}{MRR} 
& BM25  & 0.427 & 0.441 & \textbf{0.463}\\
& BM25+RM3  & 0.366 & 0.385 & \textbf{0.415}\\
& QL  & 0.402 & 0.420 & \textbf{0.454}\\
& QL+RM3  & 0.319 & 0.337 & \textbf{0.382}\\
\hline
\multirow{4}{*}{R@10} 
& BM25 & 0.636 & 0.646 & \textbf{0.679}\\
& BM25+RM3 & 0.600 & 0.617 & \textbf{0.651}\\
& QL & 0.611 & 0.633 & \textbf{0.671}\\
& QL+RM3 & 0.552 & 0.579 & \textbf{0.629}\\
\hline
\multirow{4}{*}{P@1} 
& BM25 & 0.311 & 0.324 & \textbf{0.344}\\
& BM25+RM3 & 0.248 & 0.265 & \textbf{0.291}\\
& QL & 0.289 & 0.302 & \textbf{0.334}\\
& QL+RM3 & 0.202 & 0.215 & \textbf{0.255}\\
\hline
\multirow{4}{*}{MAP} 
& BM25 & 0.422 & 0.435 & \textbf{0.457}\\
& BM25+RM3 & 0.361 & 0.380 & \textbf{0.409}\\
& QL & 0.398 & 0.414 & \textbf{0.448}\\
& QL+RM3 & 0.315 & 0.333 & \textbf{0.377}\\
\hline
\bottomrule
\Xhline{2\arrayrulewidth}
\end{tabular}
}
\end{center}
\vspace{-0.15in}
\caption{Retrieval results on MS MARCO dataset. We use following evaluation metrics: MRR, R@10, P@1 and MAP. The best performance is marked in \textbf{bold}.}
\label{tab:tab2}
\vspace{-0.20in}
\end{table}
Results on the ANTIQUE dataset and sampled MS MARCO dataset are shown in Table~\ref{tab:tab1} and Table~\ref{tab:tab2}, respectively. Our UDEG framework significantly outperforms all baselines in all evaluation metrics.  
Interestingly, the retrieval performance of QL is impressively enhanced when using our framework. Note that the retrieval performance of QL without expansion is much lower than BM25; however, QL shows comparable and even outstanding performance with our expansion framework.

\paragraph{Effectiveness of Abstractive Generation}
Compared to the extractive and the paraphrasing baselines, our proposed abstractive framework outperforms them in all metrics. 
Notably, even though PEGASUS$_{ext}$ is pre-trained on the same PEGASUS pipeline with the UDEG framework, the expansion model with the extractive generation scheme is ineffective, since it cannot solve the vocabulary mismatch problem. 
However, the proposed UDEG framework can solve it by generating novel words, which demonstrates the effectiveness of the abstractive generation scheme.

\paragraph{Effectiveness of Query Expansion}
When RM3 is applied, the performance is negatively affected in most cases. As~\citet{Nogueira2019document} reported, we can also interpret the obtained results as evidence that document expansion is more effective than query expansion since a document often contains more signals than a query with its longer length. 
\begin{figure}[t!]
\begin{center}
\includegraphics[width=0.45\textwidth]{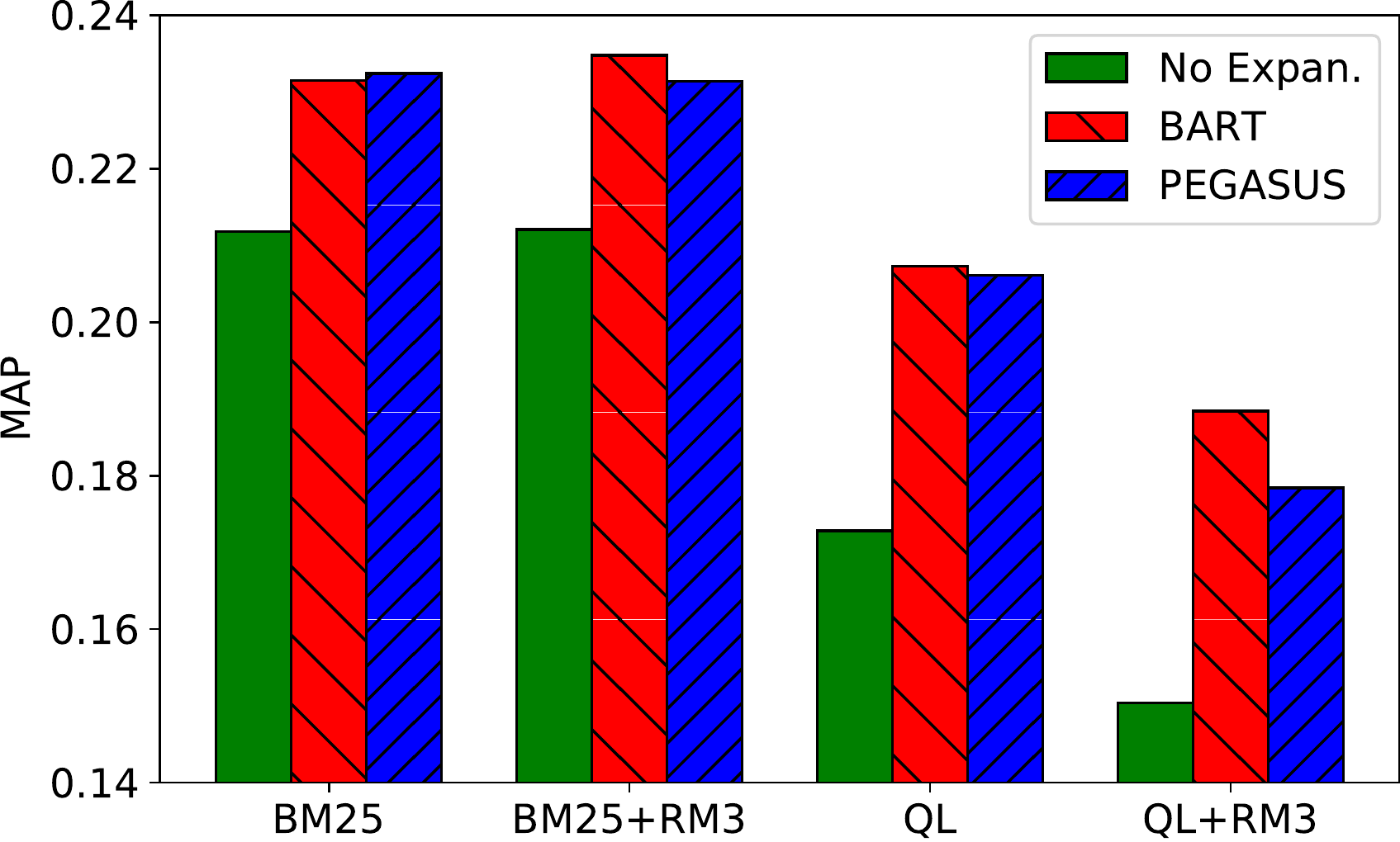}
\end{center}
\vspace{-0.15in}
\caption{
Comparison of BART and PEGASUS language models. The numbers of generated sentences for expansion are both set to one.
}
\vspace{-0.15in}
\label{fig:fig2}
\end{figure}

\subsection{Ablation and Discussion}
Which attributes contribute how much to the performance improvement? To see this, we further perform an ablation study, as follows.

\paragraph{Robustness on Different Language Models}
To validate the robustness of our framework on different language models, we compare the performances of PEGASUS and BART~\cite{Lewis2020BART}, both of which are trained on the XSUM dataset. 
As shown in Figure~\ref{fig:fig2}, the UDEG framework with PEGASUS shows performance similar to the one with BART, both of which consistently outperform the naïve baseline, which neither expands the query nor the document. Thus, the results show that the UDEG framework does not depend on a specific language model, but robustly improves the overall retrieval performance.
\begin{figure}[t]
\begin{center}
\includegraphics[width=0.4\textwidth]{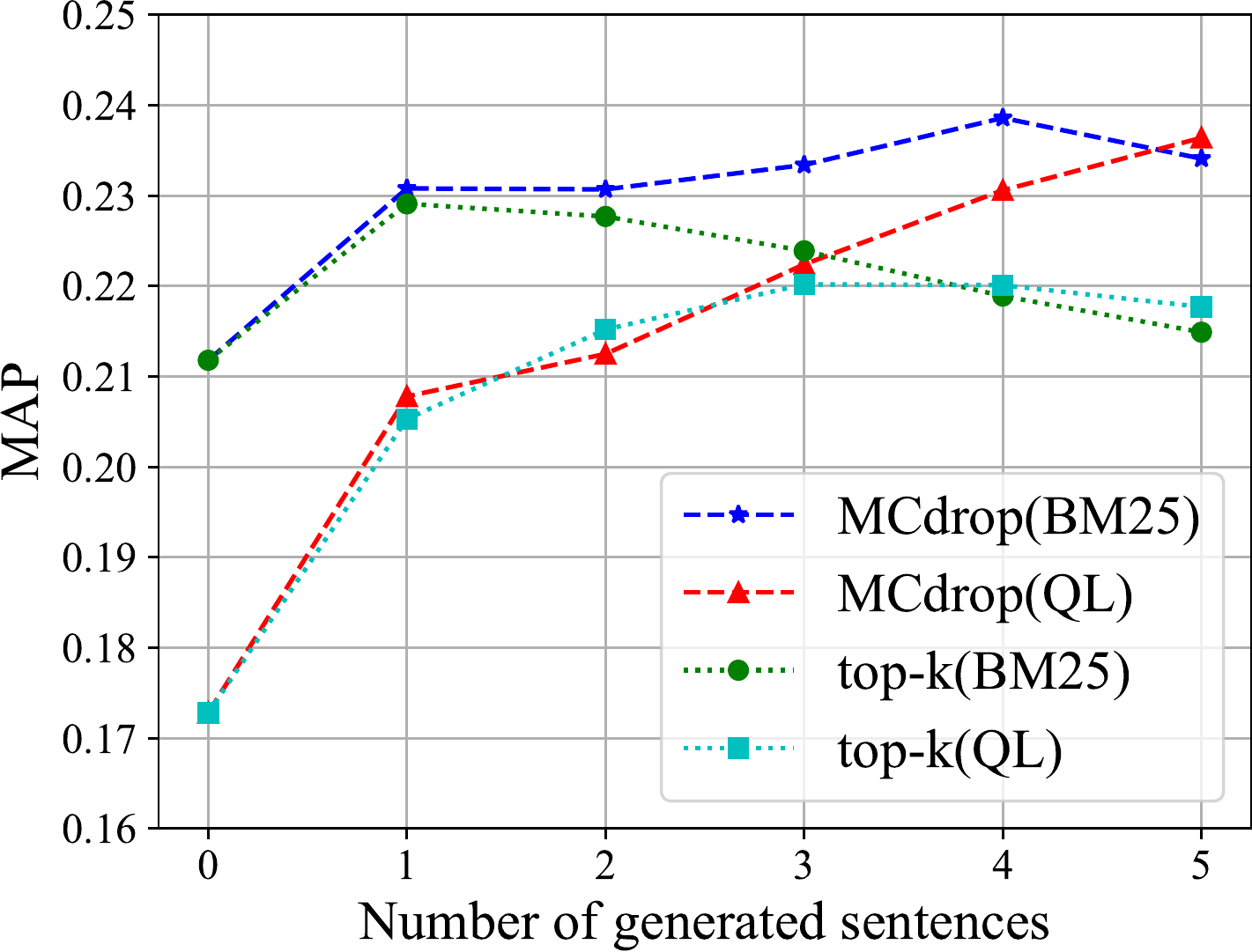}
\end{center}
\vspace{-0.15in}
\caption{
MAP scores of two different stochastic generation strategies (MC dropout vs. top-k sampling) with a varying number of generated sentences. When the number of generated sentences is 0, it refers to the naïve model without expansion.
}
\vspace{-0.1in}
\label{fig:fig3}
\end{figure}

\begin{figure}[t!]
\begin{center}
\includegraphics[width=0.4\textwidth]{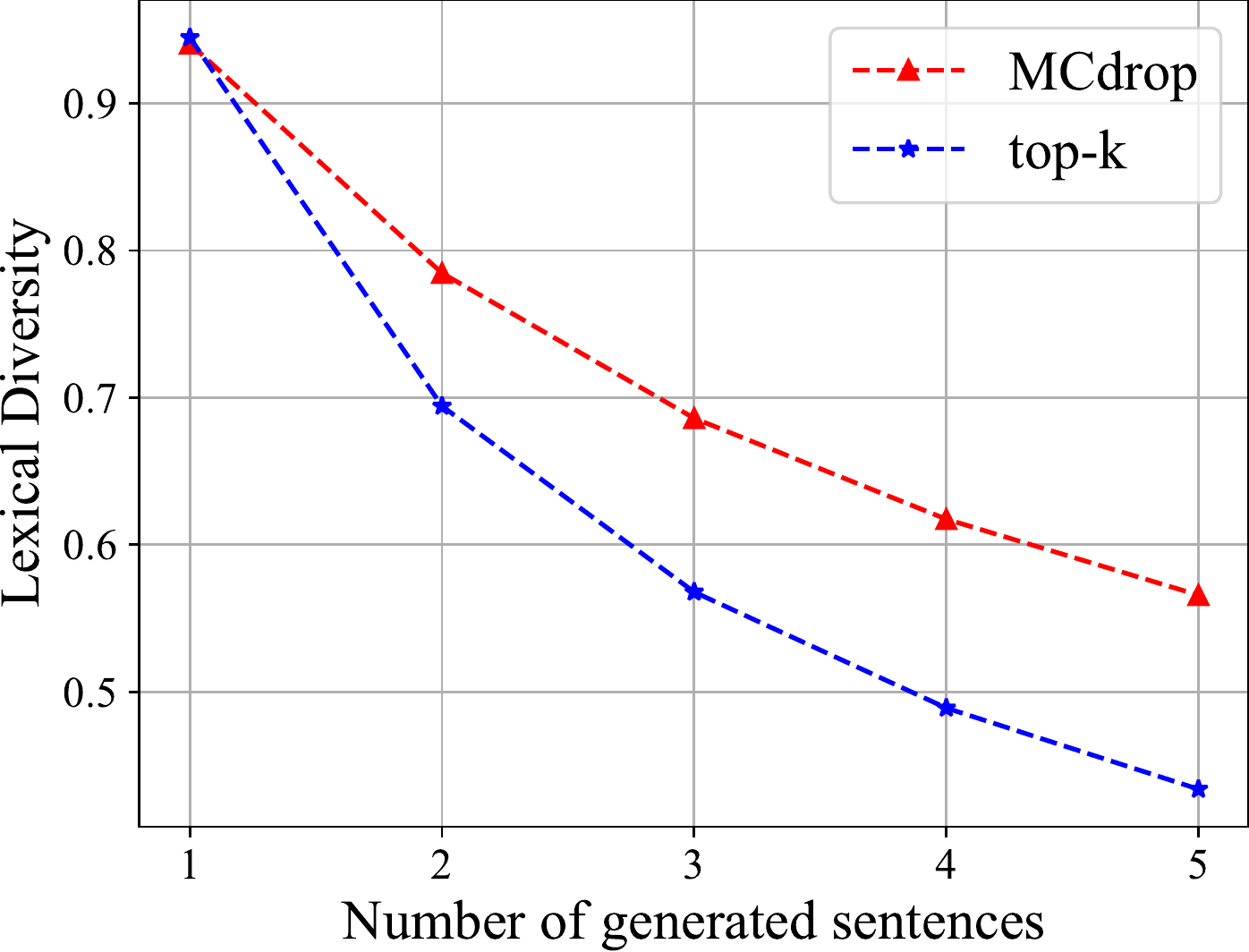}
\end{center}
\vspace{-0.15in}
\caption{
Lexical diversity of two different stochastic generation strategies (MC dropout vs. top-k sampling) with a varying number of expanded sentences.
}
\vspace{-0.2in}
\label{fig:fig4}

\end{figure}

\begin{table*}[t!]\
\small
\centering
\begin{tabular}{cl}
\toprule
\textbf{Query} & How is the chemistry is a basic of science? \\ 
\cdashline{0-1}\noalign{\vskip 0.75ex}
\textbf{\begin{tabular}[c]{@{}c@{}}Relevant\\ Document\end{tabular}} & \begin{tabular}[c]{@{}l@{}}Chemistry is a basic because all matter can be broken down into elements (i.e., hydrogen, oxygen,\\ nitrogen, etc.); without matter, nothing could be studied.	\end{tabular} \\ 
\cdashline{0-1}\noalign{\vskip 0.75ex}
\textbf{\begin{tabular}[c]{@{}c@{}}Generated\\ Sentences\end{tabular}} & \begin{tabular}[c]{@{}l@{}}\textbf{1)} Chemistry is the study of \textcolor{red}{atoms} and \textcolor{red}{molecules}. \textbf{2)} Chemistry is the study of matter and how it\\ is made. \textbf{3)} Chemistry is the \textcolor{red}{study} of matter. \textbf{4)} Chemistry is a basic \textbf{\textcolor{red}{science}}.\end{tabular} \\ 
\cdashline{0-1}\noalign{\vskip 0.75ex}
\multicolumn{2}{c}{Original Document Rank: 104 \quad\quad\quad Expanded Document Rank: 5} \\ \bottomrule

\textbf{Query} & How is the library consider as a heart of university? \\ 
\cdashline{0-1}\noalign{\vskip 0.75ex}
\textbf{\begin{tabular}[c]{@{}c@{}}Relevant\\ Document\end{tabular}} & \begin{tabular}[c]{@{}l@{}}Whatever you are studying has to be found somewhere for you to learn it. That's where the library\\ comes into focus.	\end{tabular} \\ 
\cdashline{0-1}\noalign{\vskip 0.75ex}
\textbf{\begin{tabular}[c]{@{}c@{}}Generated\\ Sentences\end{tabular}} & \begin{tabular}[c]{@{}l@{}}\textbf{1)} If you're studying at \textbf{\textcolor{red}{university}}, you'll need a library. \textbf{2)} A library is a \textcolor{red}{place} where you can \textcolor{red}{find} \\
\textcolor{red}{out} more about the \textcolor{red}{subject} you are studying. \textbf{3)} If you're studying, you'll be studying. \textbf{4)} There are\\
\textcolor{red}{many different ways} you can study.\end{tabular} \\ 
\cdashline{0-1}\noalign{\vskip 0.75ex}
\multicolumn{2}{c}{Original Document Rank: 636 \quad\quad\quad Expanded Document Rank: 32} \\ \bottomrule

\textbf{Query} & What do doctors do when a patient has a Do Not Resuscitate Order? \\ 
\cdashline{0-1}\noalign{\vskip 0.75ex}
\textbf{\begin{tabular}[c]{@{}c@{}}Relevant\\ Document\end{tabular}} & 
\begin{tabular}[c]{@{}l@{}}All healthcare professionals involved in the care of that patient will not do anything to prolong\\ the patient's life if in case patient deteriorates/dies. DNR orders may be modified, some may choose \\mechanical ventilation, or drugs. Usually when a pt is DNR, comfort measures is provided only.\end{tabular} \\ 
\cdashline{0-1}\noalign{\vskip 0.75ex}
\textbf{\begin{tabular}[c]{@{}c@{}}Generated\\ Sentences\end{tabular}} & 
\begin{tabular}[c]{@{}l@{}}\textbf{1)} DNR is not \textcolor{red}{life-support}. \textbf{2)} When a patient is in a "do not resuscitated" (DNR) state, that patient's \\life will not be \textcolor{red}{saved}. \textbf{3)} A DNR is a \textcolor{red}{decision} made by the patient's \textcolor{red}{family} or health care provider to \\prolong the life of the patient. \textbf{4)} A "do not resuscitate"(DNR) order does not \textcolor{red}{mean} that a patient \\should be put on life \textcolor{red}{support}.\end{tabular} \\ 
\cdashline{0-1}\noalign{\vskip 0.75ex}
\multicolumn{2}{c}{Original Document Rank: 40 \quad\quad\quad Expanded Document Rank: 1} \\ \bottomrule

\end{tabular}
\vspace{-0.05in}
\caption{Examples of generated sentences by the UDEG framework on the ANTIQUE dataset. Note that the first example contains scientific information.
The generated terms are highlighted in \textcolor{red}{red} if the terms are novel but relevant to the document, and further highlighted in \textbf{bold} if the novel terms appear in the query.}
\label{tab:example}
\vspace{-0.15in}
\end{table*}


\paragraph{Comparison of Stochastic Generation Strategy}
We compare two stochastic generation strategies, MC dropout and top-k sampling. 
The top-k sampling is designed to generate diverse outputs by sampling the next word from the $k$ most likely candidates, instead of deterministically selecting the next word~\cite{Fan2018topk}. 
As shown in Figure~\ref{fig:fig3}, even though both strategies aim at generating diverse sentences stochastically, the MC dropout strategy outperforms the top-k sampling strategy. 
Where does this performance difference come from? The hypothesis is that MC dropout makes more diverse terms across sentences than top-k sampling. Specifically, we often obtain the same starting words from top-k sampling, which leads to generate a number of sentences that might share same starting words. On the other hand, MC dropout randomly perturbs the embeddings at the beginning of generating each sentence, which leads to a diversity of terms even at the starting point. 
To verify this hypothesis, we compare the lexical diversity of MC dropout and top-k sampling strategies with a varying number of generated sentences. The lexical diversity is calculated by averaging the proportion of the unique unigrams in generated sentences for each document. 
As Figure~\ref{fig:fig4} shows, the lexical diversities of the generated sentences by top-k sampling are consistently lower and drop more rapidly than that by MC-dropout.

\paragraph{Varying the Number of Expanded Sentences}

To understand how stochastically generated sentences with MC dropout improves the retrieval performance, we experiment our UDEG with a varying number of generated sentences on two retrieval models, BM25 and QL.
Figure~\ref{fig:fig3} shows that the performances of both models tend to improve with increasing numbers of expanded sentences. Interestingly, QL is largely improved as stochastically generated sentences are stacked up to the original document. Meanwhile, the performance is slightly dropped when expanding five sentences for BM25. 
These results indicate that setting an appropriate number of generated sentences is important for optimal results, since too much information may degrade the context of the original document.

\subsection{Case Study}

For a qualitative analysis, we conduct a case study to explore the strengths of the UDEG framework. 
Table~\ref{tab:example} shows examples of successfully retrieved expanded-documents with the UDEG framework compared to the original documents without expansion. 
Note that the original documents are retrieved with lower ranks, but get higher ranks after applying the UDEG framework. 
We note that the generated sentences contain novel words, while they sometimes contain copied terms. 
This tendency of copying increases the importance of the keyphrases which contributes to the effective term re-weighting.
At the same time, newly generated terms are found to resolve the vocabulary mismatch problem by introducing synonyms or semantically related terms.
These findings advocate for the importance of using abstractly generated sentences for document expansion in \textit{ad-hoc} retrieval systems, which can help term re-weighting and alleviate the vocabulary mismatch problem at the same time.

\section{Conclusion}
We presented a novel framework, which we refer to as \textbf{U}nsupervised \textbf{D}ocument \textbf{E}xpansion with \textbf{G}eneration (UDEG), that generates diverse terms with stochastic perturbation over pre-trained language models, and efficiently enriches the document representation, without using any query information for training. Remarkably, UDEG employed in a retrieval system shows significant performance improvements on two standard benchmark datasets. Also, a detailed analysis shows that an abstractive generation framework with stochastic perturbation positively contributes to the retrieval performance.
Not only synonymy, but also other problems of the IR system such as polysemy could be addressed using our UDEG framework, to be left for the future work. 
We believe that the benefits of using diversely generated document-relevant sentences would allow further improvements on any IR system, targeting at scholarly and scientific information. 

\section*{Acknowledgements}
This work was supported by Institute for Information
and communications Technology Promotion
(IITP) grant funded by the Korea government
MSIT) (No. 2018-0-00582, Prediction and augmentation
of the credibility distribution via linguistic
analysis and automated evidence document collection).

\bibliography{anthology,custom}
\bibliographystyle{acl_natbib}


\end{document}